\documentclass[12pt ]{article}
\usepackage{amsmath,amsfonts,amsthm,amsbsy}
\usepackage{graphicx}
\numberwithin{equation}{section}
\def\cite#1{[\ref{#1}]}

\def\zb{\overline{z}}
\def\zb1{\overline{z}_{1}}
\def\zb2{\overline{z}_{2}}
\def\z1{z_{1}}
\def\z2{z_{2}}

\def\l{\lambda}

\newtheorem{lem}{Lemma}

\theoremstyle{definition}

\begin{document} 
\pagenumbering{roman}
\setcounter{page}{0}

\title{Proof and extension of the resistance formula for an $m\times n$ cobweb network conjectured by Tan, Zhou and Yang }
 
\author{J W Essam\dag \, Zhi-Zhong Tan\dag\dag \, and F Y Wu \ddag
     \thanks{email: {\tt j.essam@rhbnc.ac.uk, tanz@ntu.edu.cn, fywu@neu.edu}}\\
\vspace{0.05 in}\\
         \dag Department of Mathematics,
         Royal Holloway College, \\University of London,
         Egham,
         Surrey TW20 0EX,
         England.
\vspace{0.2 in}\\
\dag\dag Department of Physics,
Nantong University,\\
Nantong 226007, China
\vspace{0.2 in}\\
\ddag Department of Physics, Northeastern University,\\
Boston, MA 02115, USA}

\maketitle 
 
\begin{abstract} 
An $m\times n$ cobweb network consists of $n$ radial lines emanating from a center and connected by $m$ concentric $n-$sided polygons. A conjecture of Tan, Zhou and Yang for the resistance from center to perimeter of the cobweb is proved by extending the method used by the above authors to derive formulae for $m=1,2$ and $3$ and general $n$. The resistance of an $m\times (s+t+1)$ fan network from the apex to a point on the boundary distant $s$ from the corner is also found.
\end{abstract} 


 
\pagenumbering{arabic}

\vskip 0.25 in

\section{Introduction}
The case $m=6$, $n=8$ of an $m\times n$ cobweb network is shown in figure \ref{cobweb}. On the basis of results for general $n$ and $m=1$, $m=2$ \cite{T11} and $m=3$ \cite{TZY} it has been conjectured by Tan, Zhou and Yang \cite{TZY} that the resistance $R_{cob}$ for general $m$ between the center and perimeter of the cobweb is given by equation \eqref{R}. A proof for $m=4$ has been given by Tan,  Zhou and Lou \cite{TL}. The conjecture for general $m$ and $n$ has recently been proved by Izmailian, Kenna and Wu   \cite{IKW} as a special case of the resistance between an arbitrary pair of nodes. They use  a modification of a method due to Wu \cite{W} which determines the point to point resistance of a general network of resistors. The new method avoids the zero eigenvalue of the Kirchoff matrix by expressing the resistance in terms of the eigenvalues and eigenvectors of a principal cofactor. The method when applied to the cobweb involves two eigenvalue problems and results in a double summation. 

The proof described here is an extension of the method of Tan et al \cite{TZY} for the case $m=3$ and is more direct and consequently shorter. The solution of only one eigenvalue problem is required together with the solution of a linear recurrence relation. By changing only the boundary conditions for the recurrence relation the resistance of a fan network, which is a segment of the cobweb network (see figure \ref{fan}), is also obtained.

The conjecture for the cobweb is stated in section 2.1 and proved in section 2.2 for general $m$ and $n$. The proof uses a Lemma which is proved in the Appendix.   In section 3 the resistance of the fan network is obtained.  
\section{The $m\times n$ cobweb network}

\subsection{The conjecture}
Let $R_{cob}$ be the resistance from center to any point on the boundary of an $m\times n$ cobweb network.
\begin{equation}\label{R}
R_{cob} =\frac{r}{2m+1} \sum_{i=1}^m (2+v_i) \frac{\coth(n \ln \sqrt{\l_i})}{\l _i -\bar \l_i}
\end{equation}
where $r$ is the resistance of the transverse elements, $v_i = 2\cos[\frac{(2i-1)\pi}{ 2m+1}]$ and 
$\l_i, \bar \l_i$ are the greater and lesser solutions of
\begin{equation}\label{quad}
\l_i^2 -u_i \l_i+1=0.
\end{equation}
Here $u_i = 2+\frac r {r_0} (2-v_i)$ where $r_0$ is the resistance of the radial elements. Notice that the denominator in \eqref{R} is equal to $\sqrt{u_i^2-4}$.
\begin{figure}[htbp]
\begin{center}
\includegraphics[scale=0.6]{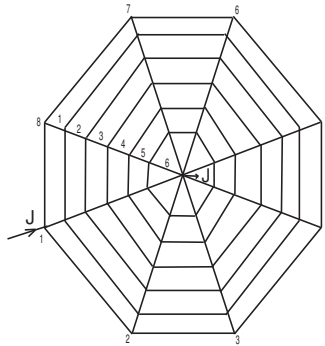}
\caption{\it A $6\times 8$ cobweb network}   \label{cobweb}
\end{center}
\end{figure}

 \subsection{The proof}
 
\begin{figure}[htbp]
\begin{center}
\includegraphics[scale=1.5]{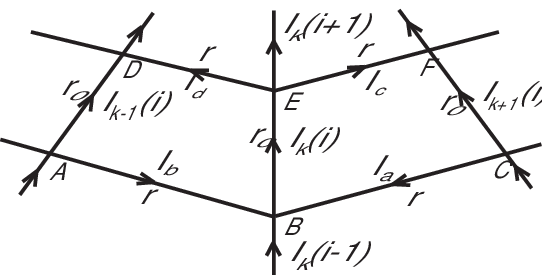}
\caption{\it The voltage loop ABEFCBEDA}   \label{loop}
\end{center}
\end{figure}

Here we extend the method used in \cite{TZY} to general $m$. Label the radial lines of the  cobweb network by $k=1,2,\dots n$.   To determine $R_{cob}$ suppose that  a current $J$ is injected into vertex $k=1$ of the perimeter and flows out from the center. Let $I_k(i)$ be the resulting current in the $i^{th}$ resistor from the edge of the $k^{th}$ radial line flowing towards the center (see figure \ref{loop}). Using Ohm's law $R_{cob}$ is given in terms of these currents by
\begin{equation}\label{Res}
R_{cob}= \frac{r_0} {J} \sum _{i=1}^m I_1(i)
\end{equation}

To determine the radial currents consider the voltage loop $ABEFCBEDA$, shown in figure \ref {loop}, centered on the $i^{th}$ resistor of the $k^{th}$ radial line .   Charge conservation gives
\begin{equation}\label{Icon}
I_a+I_b =I_k(i) -I_k(i-1)  \qquad \hbox{and} \qquad I_c+I_d= I_k(i)-I_k(i+1)  
\end{equation}
For the moment we assume that current $J_k$ is injected into perimeter vertex $k$ so that when $i=1$ in \eqref{Icon}, $I_k(i-1) = J_k$.
The sum of the voltage differences round the loop is zero so using Ohm's law
\begin{equation}
r_0(2I_k(i) - I_{k-1}(i) -I_{k+1}(i)) +r(I_a+I_b) +r_1(I_c+I_d)=0
\end{equation}
where $r_1=r$ for $i<m$ and is zero for $i=m$. Combining these equations
\begin{equation}\label{Ieq}
I_{k+1}(i) =-h I_k(i-1)+ (h+h_1+2)I_k(i)   -h_1 I_k(i+1) -I_{k-1}(i)  
\end{equation}
where $h=r/r_0$, $h_1= r_1/r_0$. 

Equation \eqref{Ieq} may be written in matrix form
\begin{equation}\label{Ivec}
I_{k+1}= [(2h+2)U_m - h V_m]I_k - I_{k-1} -h\{J_k,0,0,\dots,0\}^T 
\end{equation}
where $U_m$ is an $m$-dimensional unit matrix and 
\begin{equation}
{\tiny V_m = \left (\begin{array}{lllllllll}
0&1&0&0&\dots&0&0&0\\
1&0&1&0&\dots &0&0&0\\
0&1&0&1&\dots &0&0&0\\
\vdots&\vdots&\vdots &\vdots&\dots&\vdots &\vdots&\vdots&\\
0&0&0&0&\dots&0&1&0\\
0&0&0&0&\dots& 1&0&1\\
0&0&0&0&\dots&0&1&1
\end{array} \right )}
\end{equation}
$V_m$ has eigenvalues $v_i=2 \cos [\frac{(2i-1)\pi}{2m+1}]$ and eigenvectors $\psi_i, i=1,2,\dots,m$. The $j^{th}$ component of $\psi_i$ , normalised so that $\psi_i(1)=1$, is given by \cite{IKW}
\begin{equation}
\psi_i(j) = \frac{\sin[\frac{(2i-1)\pi j}{2m+1}]}{\sin[\frac{(2i-1)\pi }{2m+1}]}
\end{equation}
\begin{lem}
For $j=1,2,\dots m$
$$\sum_{i=1}^m (2+v_i) \psi_i(j) = 2m+1 \qquad \hbox{independently of $j$}$$
\end{lem}
\noindent The proof of this lemma is given in appendix A. Notice that $v_i= \psi_i(2)$.

Now let $\Psi$ be the matrix with $i^{th}$ row $\psi_i$ and define $X_k = \Psi I_k$.
Using the lemma
\begin{equation}
\sum_{i=1}^m(2+v_i) X_k(i) = \sum_{i=1}^m(2+v_i)\sum_{j=1}^m \psi_i(j) I_k(j) = (2m+1)\sum_{j=1}^m I_k(j)\end{equation}
which combined with \eqref{Res} gives the required resistance.
\begin {equation}\label{Rsum}
R_{cob}= \frac{r_0} {(2m+1)J} \sum _{i=1}^m (2+ v_i)X_1(i)
\end{equation}
It remains to determine $X_1(i)$.

 Multiplying \eqref{Ivec}
on the left by $\Psi$, noting that $\Psi V_m$ is diagonal with diagonal elememts $v_i$ and taking the $i^{th}$ component
\begin{equation}\label{Xeq}
X_{k+1}(i) = u_i X_k(i) -X_{k-1}(i) -hJ_k 
\end{equation}
where $u_i = 2h+2 -h v_i$.
To determine $R_{cob}$ using \eqref{Res} we must set $J_k=J \delta_{k,1}$ so for $k\ge 2$ equation \eqref{Xeq} has solution
\begin{equation}\label{Xsol}
X_k(i) = A_i \l_i^k + \bar A_i \bar \l_i ^k
\end{equation}
where $\l_i$ and $\bar \l_i$ are as defined in the conjecture and from \eqref{quad}
\begin{equation}\label{la}
 \l_i + \bar\l_i =u_i \qquad \hbox{ and }\qquad
\bar \l_i =1/\l_i.
\end{equation}

The coefficients $A_i$ and $\bar A_i$ are determined in terms of $X_1(i)$ and $X_2(i)$ by
\begin{equation}\label{Aeq}
A_i =\frac{\bar\l_i(X_2(i)-\bar\l_i X_1(i))}{\l_i-\bar \l_i} \qquad \bar A_i=\frac{-\l_i(X_2(i)-\l_i X_1(i))}{\l_i-\bar \l_i}
\end{equation}
Setting $k=1$ in \eqref{Xeq} and using symmetry about the $k=1$ radial line, which gives $X_0(i)=X_2(i)$,
\begin{equation}\label{X2}
X_2(i) = \frac 12 (\l_i+\bar \l_i) X_1(i) -\frac 12 hJ.
\end{equation}
Using \eqref{X2} to eliminate $X_2(i)$ from \eqref{Aeq}
\begin{equation}\label{Asol}
\l_i A_i =\frac 12 \left (X_1(i) -\frac{h J}{\l_i-\bar \l_i} \right ) \qquad \bar \l_i \bar A_i=\frac 12 \left (X_1(i) +\frac{hJ}{\l_i-\bar \l_i} \right )
\end{equation}
and using the periodicity $X_{n+1}(i) = X_1(i)$, with $k=n$ in \eqref{Xsol} together with \eqref{Asol}
\begin{equation}
X_1(i) = A_i\l_i^{n+1} +\bar A_i \bar \l_i^{n+1} = \frac{(\l_i^n-\bar \l_i^n)h J}{(\l_i^n+\bar \l_i^n -2)(\l_i-\bar \l_i)}
\end{equation}
The final result \eqref{R} follows by cancellation of a factor $(\sqrt \l_i)^n-(\sqrt{ \bar\l_i})^n$ and substitution in \eqref{Rsum}.

\section{Resistance of the $m\times (s+t+1)$ fan}
\begin{figure}[htbp]
\begin{center}
\includegraphics[scale=0.8]{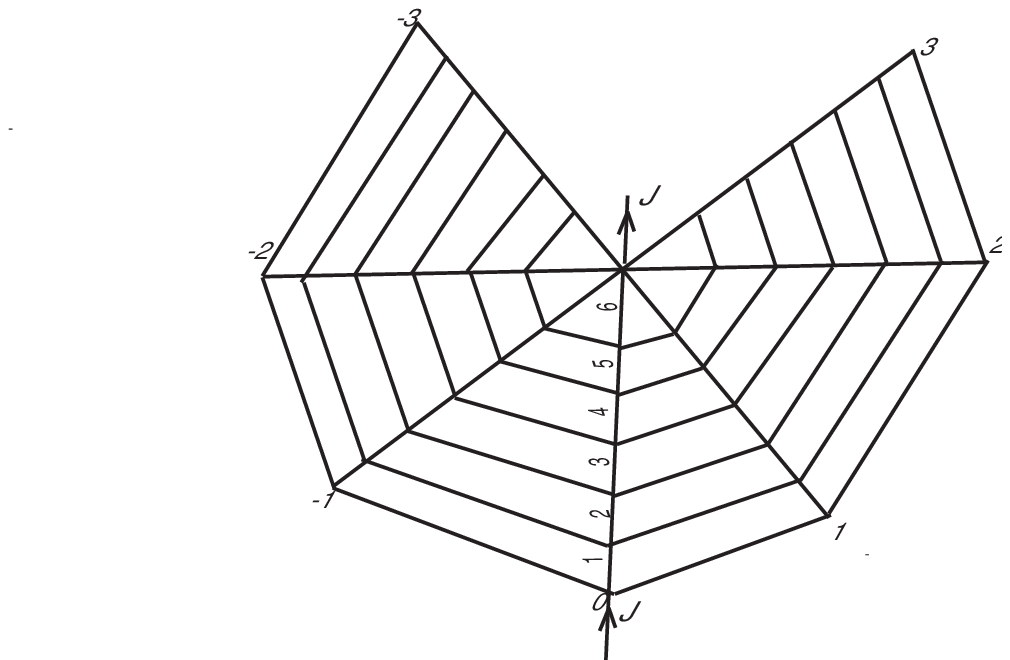}
\caption{\it The $s=t=3$ symmetric fan network}   \label{fan}
\end{center}
\end{figure}

The fan has $s+t+1$ radial lines labelled from $k=-s$, to $k=t$. It is a segment of the cobweb (see figure \ref{fan} ) and its resistance will be determined between the end of the $k=0$ radial line and the apex.

Equation \eqref{Ivec} now holds for $-s<k<t$ with $J_k=J\delta_{k,0}$. For $k=t$ we only use the loop ABEDA in figure \ref{loop} to obtain the boundary equations
\hspace{-1.0in}
\begin{align}
I_{t-1}(i) =(2h+1)I_t(i)-&hI_t(i-1)-hI_t(i+1)\qquad \hbox{ for $i<m$}\\
I_{t-1}(m) = (h+1)I_t(m) -&hI_t(m-1)\\
\intertext{or in matrix form}
I_{t-1}=((2h+1)U_m -&hV_m)I_t\\
\intertext{which on multiplying by $\Psi$ and taking the $i^{th}$ component gives}
 X_{t-1}(i) = (u_i-1)X_t(i)& \qquad \hbox{and similarly} \qquad  X_{-s+1}(i) = (u_i-1)X_{-s}(i)\label{B1}
\end{align}
In the region $k>0$ the solution \eqref{Xsol} will be used and in the $k<0$ region different coefficients are required. With $\ell =-k$
\begin{equation}\label{Xsolneg}
X_{-\ell}(i) = B_i \l_i^\ell + \bar B_i \bar \l_i ^\ell
\end{equation}
On the $k=0$ radial line where the current is injected equation \eqref{Xeq} becomes
\begin{equation}
X_{-1}(i)+X_1(i) =u_i X_0(i) -hJ\label{B2}
\end{equation}
Applying \eqref{B1} and \eqref{B2} to \eqref{Xsol} and \eqref{Xsolneg} produces the simultaneous equations
\begin{align}
\bar A_i &= \l_i^{2t+1} A_i \qquad\bar B_i = \l_i^{2s+1} B_i\label{Acon1}\\
X_0(i) &= A_i + \bar A_i = B_i+\bar B_i\\
A_i - \bar B_i &=\frac{-hJ}{\l_i-\bar\l_i}\label{Acon2}
\end{align}
Now 
\begin {equation}
R_{fan}= \frac{r_0} {(2m+1)J} \sum _{i=1}^m (2+ v_i)X_0(i) = \frac{r_0} {(2m+1)J} \sum _{i=1}^m (2+ v_i)(A_i+\bar A_i)
\end{equation}
Using \eqref{la} and solving the simultaneous equations for $A_i$ and $\bar A_i$ gives the final result
\begin{equation}\label{Rfanasym}
R_{fan}=\frac{r}{2m+1} \sum_{i=1}^m  \frac{(2+v_i) (\l_i^{s+t+1}+\bar \l_i^{s+t+1} +\l^{t-s}+\l^{s-t})}{(\l _i -\bar \l_i)(\l_i^{s+t+1}-\bar \l_i^{s+t+1}))}
\end{equation}
With $s+t+1=n$, the number of radial lines, the symmetric case $s=t$ reduces to
\begin{align}\label{Rfan}
R_{fan}^{sym}&= \frac{r}{2m+1} \sum_{i=1}^m  \frac{(2+v_i)}{\l _i -\bar \l_i }\frac{\l^\frac n 2+\bar\l^\frac n 2}{\l^\frac n 2-\bar\l^\frac n 2}\\
& =\frac{r}{2m+1} \sum_{i=1}^m  \frac{(2+v_i)}{\l _i -\bar \l_i } \coth (n\ln \sqrt{\l_i}) = R_{cob}
\end{align}
The resistance equality of the symmetric fan and cobweb networks having the same number $n$ of radial lines is at first surprising. However the initial and final radial lines of the fan may be joined to form a cobweb and by symmetry the added resistors carry no current. The current distributions in the two networks must therefore be the same.

\section{Summary and discussion.}
The method of Tan, Zhou and Yang \cite{TZY} has been used to derive formulae \eqref{R} and \eqref{Rfanasym} for the resistance of the cobweb and fan networks shown in figures \ref{cobweb} and \ref{fan}. In the case of the cobweb network the formula is for the resistance between the center and any point on the boundary. For the fan network the resistance is obtained between the apex and a point on the boundary distant $s$ from one corner and $t$ from the other. In the symmetric case $s=t$, shown in the figure, the resistance is the same as for a cobweb with $s+t+1$ radial lines.

 The cobweb formula was conjectured in \cite{TZY}. The conjecture has recently been proved  \cite{IKW} by a method which gives the general point-to-point resistance in terms of a double summation.  Significant further analysis is then required to reduce this double summation to a single summation and hence to \eqref{R}. The method also requires the solution of two eigenvalue problems.

By contrast the method of Tan et al splits the derivation into two parts. The first part solves the radial equation and involves an eigenvalue problem. The second part uses a Lemma which applies to both networks and states a surpising relation between the eigenvectors. This Lemma enables the resistance to be expressed as a single sum the summand of which involves only the solution of a second order recurrence relation.  The same recurrence relation also applies to both networks, only the boundary conditions are different.

 \vspace{0.5 in}

\section{\bf \Large Appendix: Proof of Lemma 1}

With $M=2m+1$ let 
\begin{align}
f_j& \equiv\sum_{i=1}^m (2+v_i)\psi_i(j) = 2\sum_{i=1}^m\cot [\frac{(2i-1)\pi}{2M}]\sin[\frac{(2i-1)j\pi}{M}]\\
\intertext{The following sum will be needed below. For $j=1,2,\dots,2m$}
S_j &\equiv 2 \sum_{i=1}^m \cos [\frac{(2i-1)j\pi}{M}] =  \frac {\sin[\frac {2mj \pi}M]} {\sin[\frac{ j\pi} M]}=(-1)^{j+1}
\intertext{The sum has been derived using complex exponentials.  Thus for $j=1$}
f_1&=2 \sum_{i=1}^m (1+\cos[\frac{(2i-1)\pi}{M}])=2m+S_1=2m+1\\
\intertext{ and for $j=1,2,\dots,2m-1$}
f_{j+1}-f_j &=2\sum_{i=1}^m \left (\cos[\frac{(2i-1)j\pi}{M}] +\cos[\frac{(2i-1)(j+1)\pi}{M}]\right )\\
&=S_j+S_{j+1}=(-1)^{j+1} +(-1)^j=0
\end{align}
Hence $f_j =2m+1$ for $j=1,2,\dots,2m$.

\vspace{0.2 in}

{\bf \Large References.}

\begin{enumerate}
\item\label{T11} Tan Z-Z, "Resistance Network Model", (Xi'an : Xidian University Press) pp9-216 (2011)
\item\label{TZY}
Tan Z-Z, Zhou L and Yang J-H,  "The equivalent resistance of $3 \times n$ cobweb network and its conjecture of an $m \times n$ cobweb network", J Phys A: Math. Theor. {\bf 46} 195202 (2013)
\item\label{TL}
Tan Z-Z, Zhou L and Luo D-F, "Resistance and capacitance of a $4\times n$ cobweb network and two conjectures", Int. J. Circ. Theor. Appl, to appear DOI: 10. 1002/cta. 1943. (2013)
\item
\label{IKW}
Izmailian N Sh, Kenna R and Wu F Y, "The two point resistance of a resistor network: A new formulation and application to the cobweb network", arXiv:1310.1335v2 [cond-mat.stat-mech]  21 Oct 2013
\item\label{W} Wu F Y, "Theory of resistor networks: the two-point resistance", J. Phys. A: Math. Gen. {bf 37} 6653 (2004).

\end{enumerate}
\end{document}